\documentclass[preprint]{aastex}

\newcommand{\et}{et~al.\ }
\newcommand{\Ha}{H$\alpha$}
\newcommand{\Hb}{H$\beta$}
\newcommand{\HeI}{\ion{He}{1}}
\newcommand{\OIII}{[\ion{O}{3}]}
\newcommand{\NII}{[\ion{N}{2}]}
\newcommand{\HII}{\ion{H}{2}}
\newcommand{\Msun}{\ensuremath{M_\odot}}

\shorttitle{POLARIZED BROAD EMISSION LINES IN SEYFERT 2 GALAXIES}
\shortauthors{MORAN ET AL.}

\begin{document}

\title{The Frequency of Polarized Broad Emission Lines in
       Type 2 Seyfert Galaxies}

\author{Edward C.\ Moran\altaffilmark{1,2}, Aaron J.\ Barth\altaffilmark{3},
Laura E.\ Kay\altaffilmark{4,5}, and Alexei V.\ Filippenko\altaffilmark{2}}

\altaffiltext{1}{Chandra Fellow.}

\altaffiltext{2}{Department of Astronomy, University of California,
                 Berkeley, CA 94720-3411.}

\altaffiltext{3}{Harvard-Smithsonian Center for Astrophysics, 60 Garden
                 Street, Cambridge, MA 02138.}

\altaffiltext{4}{Department of Physics and Astronomy, Barnard College,
                 Columbia University, New York, NY 10027.}

\altaffiltext{5}{Visiting Astronomer, Cerro Tololo Inter-American Observatory.
                 CTIO is operated by AURA, Inc., under contract with the
                 National Science Foundation.}

\begin{abstract}

We have discovered polarized broad emission lines in five type~2 Seyfert
galaxies (NGC 424, NGC 591, NGC 2273, NGC 3081, and NGC 4507), establishing
that these objects are type~1 Seyferts obscured by dense circumnuclear
material.  The galaxies are part of a distance-limited sample of 31
Seyfert~2s, for which spectropolarimetric observations are now complete.
Combined with published reports, our results indicate that at least 11 of the
galaxies in this sample, or $\ge35$\%, possess hidden broad-line regions.
This represents the first reliable estimate of the frequency of polarized
broad emission lines in type~2 Seyferts, which has important implications
for the general applicability of Seyfert unification models.

\end{abstract}


\newpage
\section{Introduction}

Seyfert galaxies have traditionally been classified into two spectroscopic
groups based on the presence (type~1) or absence (type~2) of permitted
optical emission lines that are broader than their forbidden lines.  The
discovery of polarized broad lines in NGC~1068 (Antonucci \& Miller 1985)
and a handful of other type~2 Seyfert galaxies (Miller \& Goodrich 1990;
Tran, Miller, \& Kay 1992; Tran 1995a; Young \et 1996; Heisler, Lumsden,
\& Bailey 1997; Kay \& Moran 1998) has therefore had significant impact
on our understanding of the relationship between the two types of objects
and, more generally, of the parsec-scale geometry of Seyfert nuclei.  The
polarization, interpreted to be due to nuclear light that is scattered
into our line of sight, establishes that some Seyfert~2s harbor normal
Seyfert~1 nuclei whose innermost regions are obscured from our direct
view, presumably by dense circumnuclear material. This material may be
present in type~1 Seyferts as well, but oriented such that it does not
lie between us and the central engine of the active nucleus.

One hindrance to the extrapolation of this unified picture to {\it all\/}
Seyfert galaxies is the fact that direct evidence for obscured type~1 nuclei
in type~2 objects has been limited to relatively few examples.  Thus, to
explore the nature of the activity in Seyfert~2 galaxies as a class, we have
initiated a multiwavelength survey of a distance-limited sample of 31 objects,
beginning with spectropolarimetric observations.  In this {\it Letter}, we
announce the discovery of ``new'' hidden broad-line regions in five Seyfert~2
galaxies from this sample; full results of our survey will be presented in
a forthcoming paper (Kay \et 2000).  Because of the unbiased nature of the
survey, which is now complete, we are able to make the first reliable
assessment of the frequency of polarized broad emission lines in type~2
Seyferts.

\section{Observations and Results}

Our survey is based on the distance-limited sample of Ulvestad \& Wilson
(1989, hereafter UW89), which, at the time of its definition, comprised all
known Seyfert galaxies (31 of type~2)\footnote{UW89 considered 35 galaxies
to be Seyfert~2s (see their Table~6), but recent spectroscopy has revealed
that three of them (MS 0942.8+0950, NGC~1365, and Mrk~1126) are actually
intermediate type~1 objects.  Over the course of our survey, we have found
that one other (Mrk~745) is a high-ionization \HII\ galaxy rather than a
Seyfert.}
with recession velocities $cz <
4600$ km~s$^{-1}$ and declinations $\delta > -45^{\circ}$.  Although some
additional Seyfert~2s have been identified within the UW89 distance limit
over the past decade, most of them are inconspicuous, low-luminosity objects
that were classified either by inference, as in the instance of NGC~3147
(Moran, Halpern, \& Helfand 1994; Ptak \et 1996), or after careful starlight
template subtraction of the optical spectrum (Ho, Filippenko, \& Sargent 1997).
Thus, the UW89 sample represents a reasonably complete, distance-limited set
of {\it classical\/} Seyfert~2s.  The sample was originally used to examine
the radio properties of Seyfert galaxies; radio results obtained for the
magnitude-limited CfA sample of Seyferts (Kukula \et 1995) are consistent
with those of UW89, suggesting that the UW89 sample is free of significant
selection biases.

The spectropolarimetry component of our Seyfert~2 survey is now complete.
Observations of the five UW89 objects in which we have discovered polarized
broad emission lines were carried out at the Keck-II 10~m telescope
on (UT) 7 March 1998 and 6 January 1999, and at the CTIO 4~m telescope on
23 June 1998 and 20 June 1999.  Table~1 lists these galaxies and their
recession velocities, as well as the observatory and exposure times
employed.  Reduction and analysis of the data were performed as described
by Barth, Filippenko, \& Moran (1999) and Kay \et (1999).  The observed
continuum polarization $P_{\rm obs}$ and linear polarization position angle
(P.A.)\ $\theta$ of each object in the 5400--5600~\AA\ (rest) wavelength
range are also listed in Table~1.  For all of the galaxies, $\theta$ remains
approximately constant across the Balmer-line profiles.

Spectropolarimetric results are shown for all five galaxies in Figures 1--5.
As indicated in the upper panel in each of the Figures, only narrow emission
lines are present in the total-flux spectra of the objects.  The polarization
spectra (middle panel) of the galaxies all display similar characteristics:
low but nonzero continuum polarization, depressions in polarization at
the locations of strong narrow emission lines (e.g., \OIII\ $\lambda\lambda
4959,5007$), and increases in polarization in the wings of the \Ha\ and
(in most cases) \Hb\ lines.  The product of the polarization and
total-flux specta (bottom panel) reveals broad permitted emission lines
in the polarized flux of each galaxy.

{\it NGC 424.}  As Figure~1 indicates, there is a distinct increase in the
polarization of this object near \Ha .  The \Ha\ line in polarized flux
is very broad, with a full-width near zero-intensity (FWZI) of at least 
12,000 km~s$^{-1}$, which is much broader than the FWZI of the narrow
\NII\ + \Ha\ blend in the total-flux spectrum shown in the upper panel
of Figure~1.  There is clearly a broad component of \Hb\ in the polarized
flux spectrum as well.

{\it NGC 591 (Mrk 1157).}  Enhancements in polarization near \Ha\ and \Hb\
are subtle in this object, but as Figure~2 shows, both lines exhibit broad
components in polarized flux.  The FWZI of the broad \Ha\ line is $\sim
4000$ km~s$^{-1}$.
Following Barth \et (1999), we have assessed the significance of our
detection of \Ha\ polarization by comparing the mean values of the Stokes
parameters $Q$ and $U$ across the broad \Ha\ profile to those in adjacent
line-free regions.  We find that $Q$ and $U$ across \Ha\ deviate
from the continuum levels by 1.9~$\sigma$ and 9.5~$\sigma$, respectively.
Polarized broad \Ha\ emission was not detected by Miller \& Goodrich (1990)
in shallower observations with the Lick 3~m telescope.

{\it NGC 2273 (Mrk 620).}  This is another weakly polarized object, but
there is an obvious bump at \Ha\ in the polarization spectrum (Fig.~3),
which is more prominent in polarized flux.  The FWZI of the broad \Ha\
line is $\sim 5000$ km~s$^{-1}$.  The significance of our detection of
a broad \Ha\ feature is 8.0~$\sigma$ in Stokes $Q$, and 2.3~$\sigma$
in Stokes $U$.

{\it NGC 3081.}  Very strong enhancements of the polarization near both \Ha\
and \Hb\ are observed in this object (Fig.~4), yielding a spectacular
type~1 spectrum in polarized light.  The broad component of \Ha\ has FWZI
$\approx 7000$ km~s$^{-1}$.  It appears that the \HeI\ $\lambda 5876$
line is polarized as well.

{\it NGC 4507.}  Similar to NGC 3081, this galaxy exhibits significant
polarization enhancements near the Balmer lines, particularly \Ha\ (Fig.~5).
A bump near \HeI\ $\lambda 5876$ also appears in the polarization spectrum.
In polarized flux, the broad \Ha\ line is strong and asymmetric, with FWZI
$\approx$ 10,000 km~s$^{-1}$.

Many Seyfert~2s with polarized broad emission lines have morphologically
``linear'' radio sources in their nuclei.  In such cases, the major axis of
the radio emission is often orthogonal to the polarization position angle
$\theta$ (Antonucci 1983, 1993).  Only two objects studied here, NGC 591 and
NGC 2273, have linear radio sources; in the others, the emission is unresolved
or only marginally resolved on $\sim 1''$ scales (Ulvestad \& Wilson 1984,
1989; Morganti \et 1999).  For NGC 591, the radio axis lies along P.A.\ =
$153^{\circ}$, which is almost perpendicular to $\theta$ (Table~1).  The
radio axis of NGC 2273 lies along P.A.\ = $90^{\circ}$, whereas $\theta =
25^{\circ}$.  Because of the weak overall polarization of this object, an
interstellar component of polarization, which we have not corrected for,
could be affecting both $P_{\rm obs}$ and $\theta$.

Unpolarized starlight from the host galaxies contributes a major, if not
dominant, fraction of the observed continuum fluxes, diminishing the
polarization of the nuclear emission.  Thus, following the methods of Tran
(1995a), we have attempted to correct for the effects of starlight dilution
to obtain a better measurement of the true level of continuum polarization
in each object.  This involves decomposition of the observed spectrum into
stellar and nonstellar components using spectra of elliptical galaxies and
spiral galaxy bulges free of emission lines as starlight templates.  Best
results for our galaxies were generally obtained with spectra of NGC 224
(M31) or NGC 7457 as templates (a full description of the analysis will be
presented by Kay \et 2000).  The derived fractions of galactic starlight
$F_{\rm g}$ present in our Seyfert~2 spectra are listed in Table~1.
Corrected values of the continuum polarization $P_{\rm cor}$
[= $P_{\rm obs} / (1 - F_{\rm g})$] are also listed in the Table.

\section{Location and Geometry of the Obscuring Medium}

The currently popular picture for the parsec-scale geometry of Seyfert nuclei,
originally proposed by Antonucci \& Miller (1985), contends that the broad-line
region (BLR) and optical/UV continuum source are surrounded by dense material
that has a roughly toroidal geometry; the orientation of this torus to
our line of sight thus determines the optical spectroscopic classification of
a given Seyfert galaxy.  In some type~2 Seyferts, where the torus is presumably
edge-on, free electrons or dust grains located above the opening of the torus
scatter nonstellar continuum and broad-line photons into our line of sight.
Because the scattered emission is polarized, spectropolarimetric observations
can provide a periscopic view of the innermost regions of these objects.
Unfortunately, selection effects have often confounded studies intended to
test the general applicability of the torus model (Antonucci 1993), and it
has been unclear what fraction of the Seyfert~2 population it describes.  In
fact, a recent imaging survey by Malkan, Gorijan, \& Tam (1998) has revealed
that type~2 Seyferts are more likely than Seyfert~1s to have dust lanes and
patches close (in projection) to the nucleus, suggesting that interstellar
dust hundreds of parsecs from the active nucleus, rather than a parsec-scale
circumnuclear torus, may be the main source of extinction in the majority of
Seyfert~2 galaxies.

Spectropolarimetry provides independent evidence regarding the location and
geometry of the obscuring medium in Seyfert~2 galaxies.  As Figures 1--5
and the results of Tran (1995b) indicate, the narrow emission lines of
objects with hidden BLRs are less polarized than the continuum and broad
Balmer lines. Thus, most of the scattering takes place interior to the
narrow-line region, which, for reasonable values of the black-hole mass
($10^8\, \Msun$) and narrow emission-line velocity widths (500 km~s$^{-1}$
FWHM), has an inner dimension of order 10~pc.  For us to detect polarized
emission, the projected size of the obscuring material must be comparable
to or smaller than the size of the scattering region, implying that the
obscuration occurs close to the black hole.  In principle, significant
polarization could result from scattering of an isotropic radiation field
by a patchy medium, but the relationship between the polarization and
radio-source position angles in Seyfert~2s would then have to be a
coincidence.  The alternative is that the radiation field somehow becomes
anisotropic prior to scattering (Antonucci 1984).  In this scenario, the
high (5--16\%) starlight-corrected polarizations of NGC 3081, NGC 4507,
and several galaxies in the Tran (1995a,b) study are possible if the obscuring
material covers much of the nucleus in the equatorial plane.  We conclude,
therefore, that the material primarily responsible for the obscuration
of those Seyfert~2s with hidden BLRs and high intrinsic polarizations is
located close to the black hole and, qualitatively, is cylindrically
symmetric---a torus, for all practical purposes.   (Note that a highly
warped, thin accretion disk might also be capable of providing the
necessary scattering geometry; see Phinney 1989.) There is no reason to
suspect that the hidden-BLR Seyfert~2s lacking high starlight-corrected
polarizations are fundamentally different.  As discussed by Miller \&
Goodrich (1990), the scattering medium may be located within the mouth
of the torus, rather than above it; polarized emission would then only
be detectable in objects where the torus is tipped toward us, and in such
cases, polarization levels would be modest.  Infrared and hard X-ray
observations appear to support this picture (Heisler \et 1997; Risaliti
\et 2000).

\section{Frequency of Polarized Broad Lines in Type~2 Seyferts}

While previous spectropolarimetric observations have demonstrated that some
type~2 Seyferts are type~1 objects obscured by dense circumnuclear material,
it has been unclear what fraction of Seyfert~2s can be described this way.
A determination of the frequency of polarized broad emission lines in
Seyfert~2s, now possible with the completion of our survey of the UW89
sample, provides an important lower limit on this fraction.

Five of the 31 galaxies in the UW89 sample were known to possess hidden BLRs
prior to the start of our survey: NGC 1068 (Antonucci \& Miller 1985), Mrk~3
and
Mrk 348 (Miller \& Goodrich 1990), NGC 4388 (Shields \& Filippenko 1988; Young
\et 1996), and IC 3639 (Heisler \et 1997).  As discussed by Kay \et (2000), we
have confirmed the presence of hidden BLRs in these last two objects.  Combined
with our discovery of polarized broad emission lines in NGC~788 (Kay \& Moran
1998), the five detections presented here in NGC 424, NGC 591, NGC 2273, NGC
3081, and NGC 4507 have more than doubled the number of known hidden BLRs in
this sample, bringing the total to 11.  Higher quality observations could,
of course, increase this number.  Thus, at least 35\% of nearby classical
Seyfert~2s have polarized broad permitted emission lines, indicating that
they are
obscured primarily by material located a few parsecs from the nucleus, rather
than by large-scale foreground dust lanes.  The remainder, however, may be
obscured mainly by interstellar matter, as suggested by Malkan \et (1998),
or they may simply lack BLRs.

Hard X-ray observations of Seyfert~2 galaxies complement spectropolarimetric
data by providing a measure of the amount of absorbing material present.
For example, recent hard X-ray observations have shown that $\sim$~50\% of
Seyfert~2s (from a different sample) have absorption column densities too high
($N_{\rm H} > 10^{24}$ atoms~cm$^{-2}$) to be associated with interstellar
dust features (Risaliti, Maiolino, \& Salvati 1999).  This is consistent with
the lower limit of 35\% we have derived for the fraction of Seyfert~2s that
are obscured by circumnuclear tori.  Nearly all of the Seyfert~2s in the
UW89 sample have been observed in the hard X-ray band, and in a future paper
we will combine those observations with our spectropolarimetric results to
explore further the nature of the obscuring material in type~2 Seyfert
galaxies.

\acknowledgments

We are very grateful to Michael Eracleous for assistance with the starlight
fraction measurements, and to A.~M.\ Magalh\~aes for supplying the polarimetry
optics used at CTIO.  The W.~M.~Keck Observatory is operated as a scientific
partnership among the California Institute of Technology, the University of
California, and NASA, and was made possible by the generous financial support
of the W.~M.\ Keck Foundation.  E.~C.~M.\ is supported by NASA through
Chandra Fellowship grant PF8-10004 awarded by the Chandra X-ray Center, which
is operated by the Smithsonian Astrophysical Observatory for NASA under
contract NAS 8-39073.  The work of A.~J.~B.\ is supported by a postdoctoral
fellowship from the Harvard-Smithsonian Center for Astrophysics.
L.~E.~K.\ acknowledges the support of the NSF through CAREER grant
AST-9501835.  We also acknowledge NASA grant NAG 5-3556.

\begin{center}
\begin{deluxetable}{lccccccc}
\tablewidth{0pt}
\tablecaption{Spectropolarimetric Observations and Results}
\tablehead{\colhead{Galaxy} &
           \colhead{$cz$ (km s$^{-1}$)} &
           \colhead{Obs.} &
           \colhead{Exp.\ (s)} &
           \colhead{$P_{\rm obs}$ (\%)} &
           \colhead{$\theta$ (deg)} &
           \colhead{$F_{\rm g}$} &
           \colhead{$P_{\rm cor}$ (\%)}}
\startdata
NGC 424  & 3496 & CTIO & 3600 & $2.31 \pm 0.03$ & $43 \pm 1$ & 0.45 & ~4.2 \\
NGC 591  & 4547 & Keck & 1200 & $0.53 \pm 0.07$ & $70 \pm 4$ & 0.52 & ~1.1 \\
NGC 2273 & 1871 & Keck & 1200 & $0.26 \pm 0.07$ & $25 \pm 7$ & 0.80 & ~1.3 \\
NGC 3081 & 2385 & Keck & 1200 & $1.43 \pm 0.07$ & $79 \pm 1$ & 0.89 & 13.0 \\
NGC 4507 & 3538 & Keck & ~800 & $0.79 \pm 0.05$ & $37 \pm 2$ & 0.87 & ~6.1
\enddata
\tablecomments{$P_{\rm obs}$, $\theta$, $F_{\rm g}$, and $P_{\rm cor}$ are
continuum values measured in the 5400--5600 \AA\ (rest) wavelength range.}
\end{deluxetable}
\end{center}

\clearpage

\begin{figure}
\epsscale{0.7}
\plotone{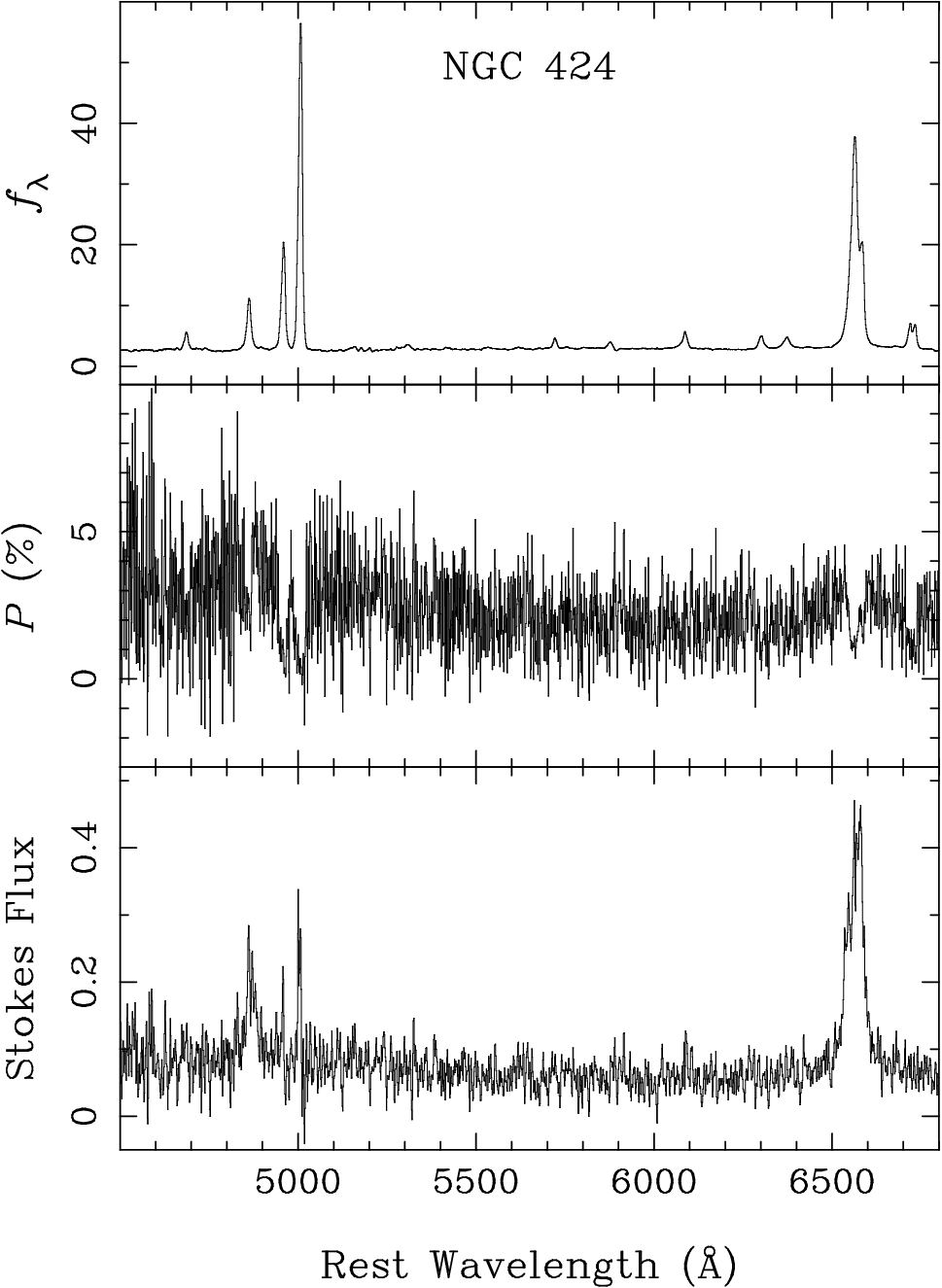}
\caption{Spectropolarimetry of NGC 424.  Corrections for
interstellar polarization and starlight dilution have not been applied.
Only narrow emission lines are present in the total-flux spectrum of the
galaxy ({\it top panel}), which has units of $10^{-15}$ erg cm$^{-2}$ s$^{-1}$
\AA$^{-1}$.  The polarization spectrum ({\it middle panel}) shows that the
polarization is enhanced in the wings of the \Ha\ and \Hb\ lines.  (The
polarization $P$ shown here is actually the ``rotated Stokes parameter'';
see Miller, Robinson, \& Goodrich 1988.)  The polarized flux, or ``Stokes
flux'' ({\it bottom panel}) is the product of the total-flux and
polarization spectra.  Broad components of both \Ha\ (FWZI $\approx$ 12,000
km~s$^{-1}$) and \Hb\ are evident in polarized flux.}
\end{figure}

\begin{figure}
\epsscale{0.7}
\plotone{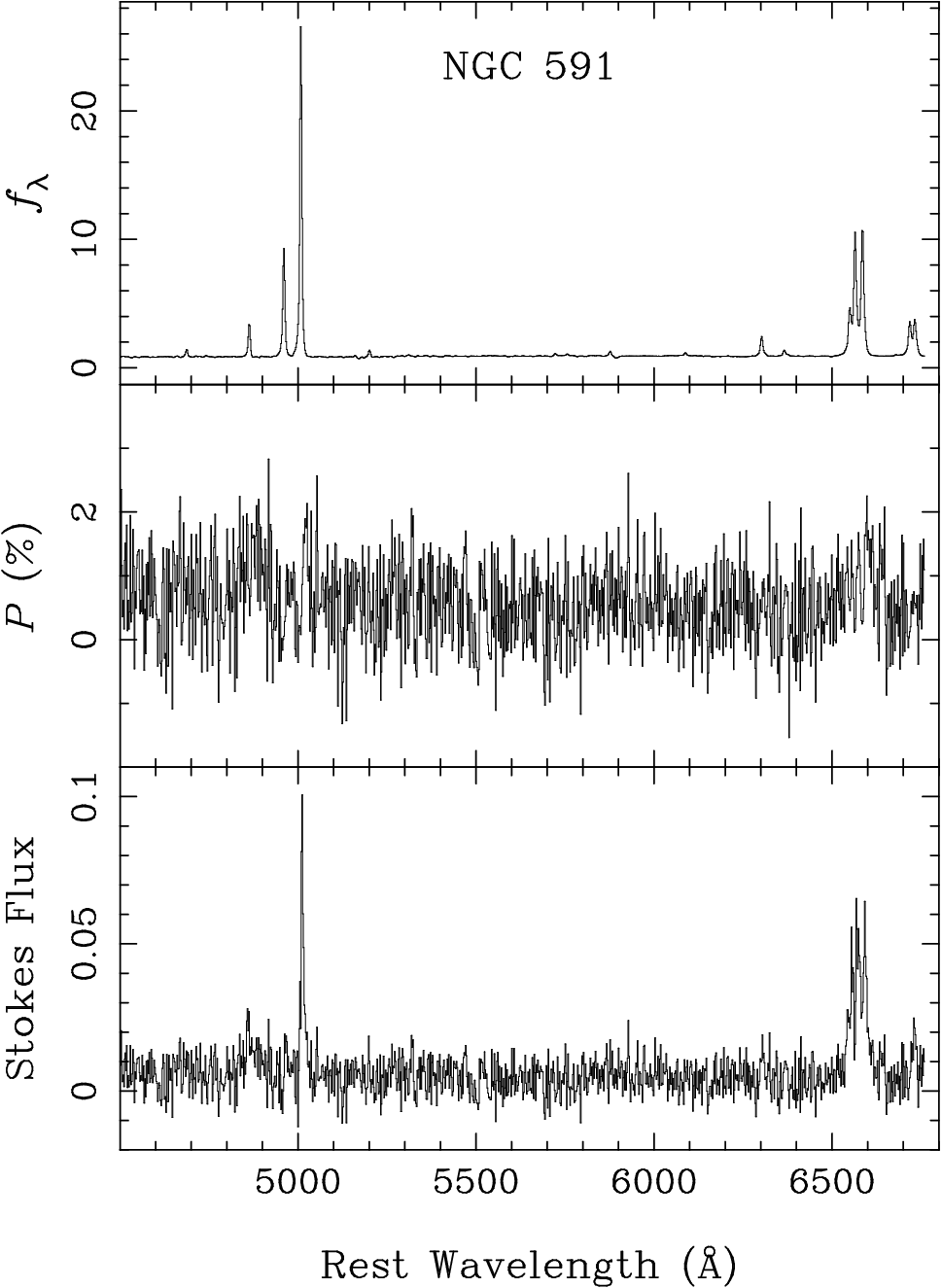}
\caption{Spectropolarimetry of NGC 591.}
\end{figure}

\begin{figure}
\epsscale{0.7}
\plotone{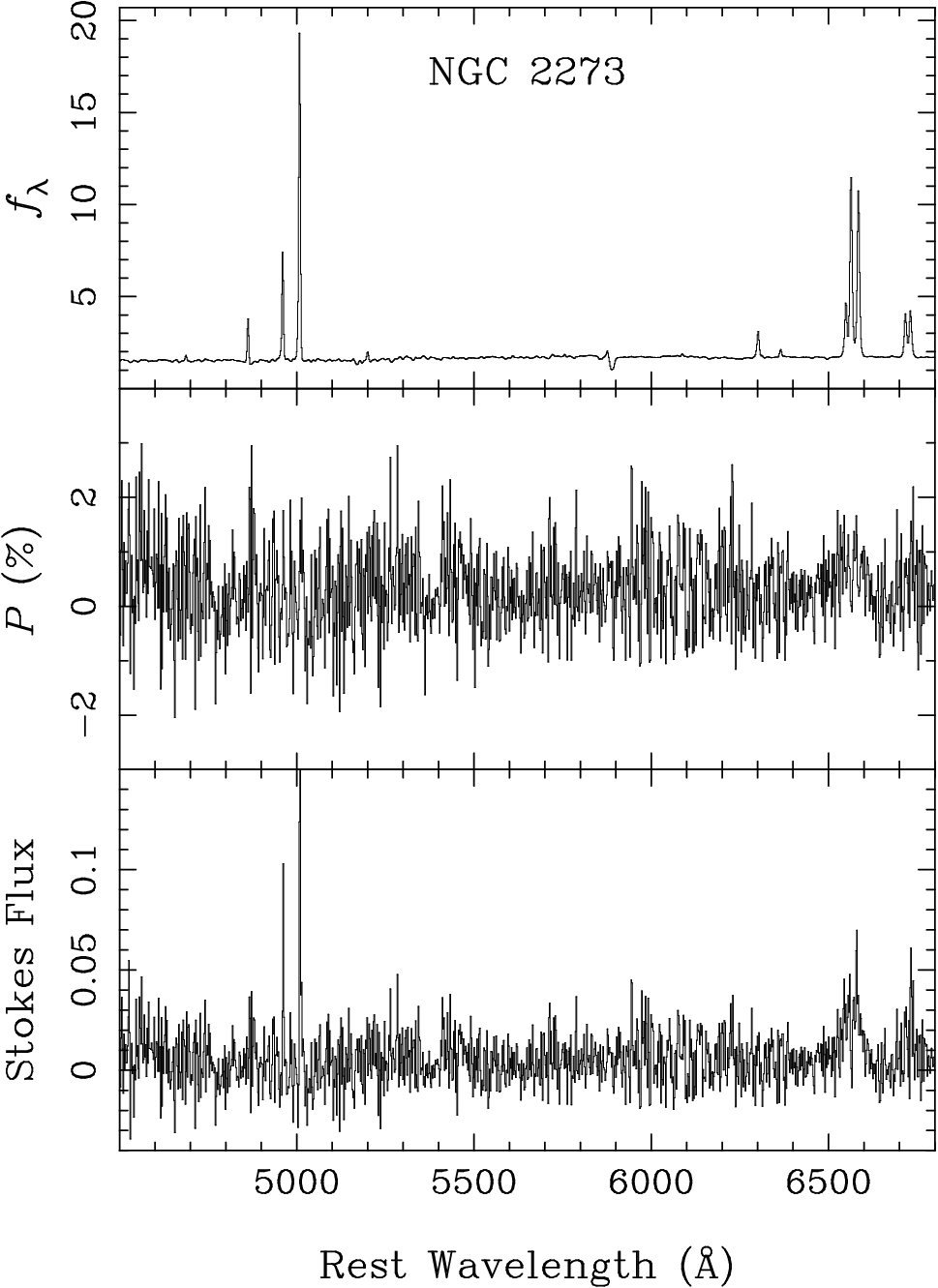}
\caption{Spectropolarimetry of NGC 2273.}
\end{figure}

\begin{figure}
\epsscale{0.7}
\plotone{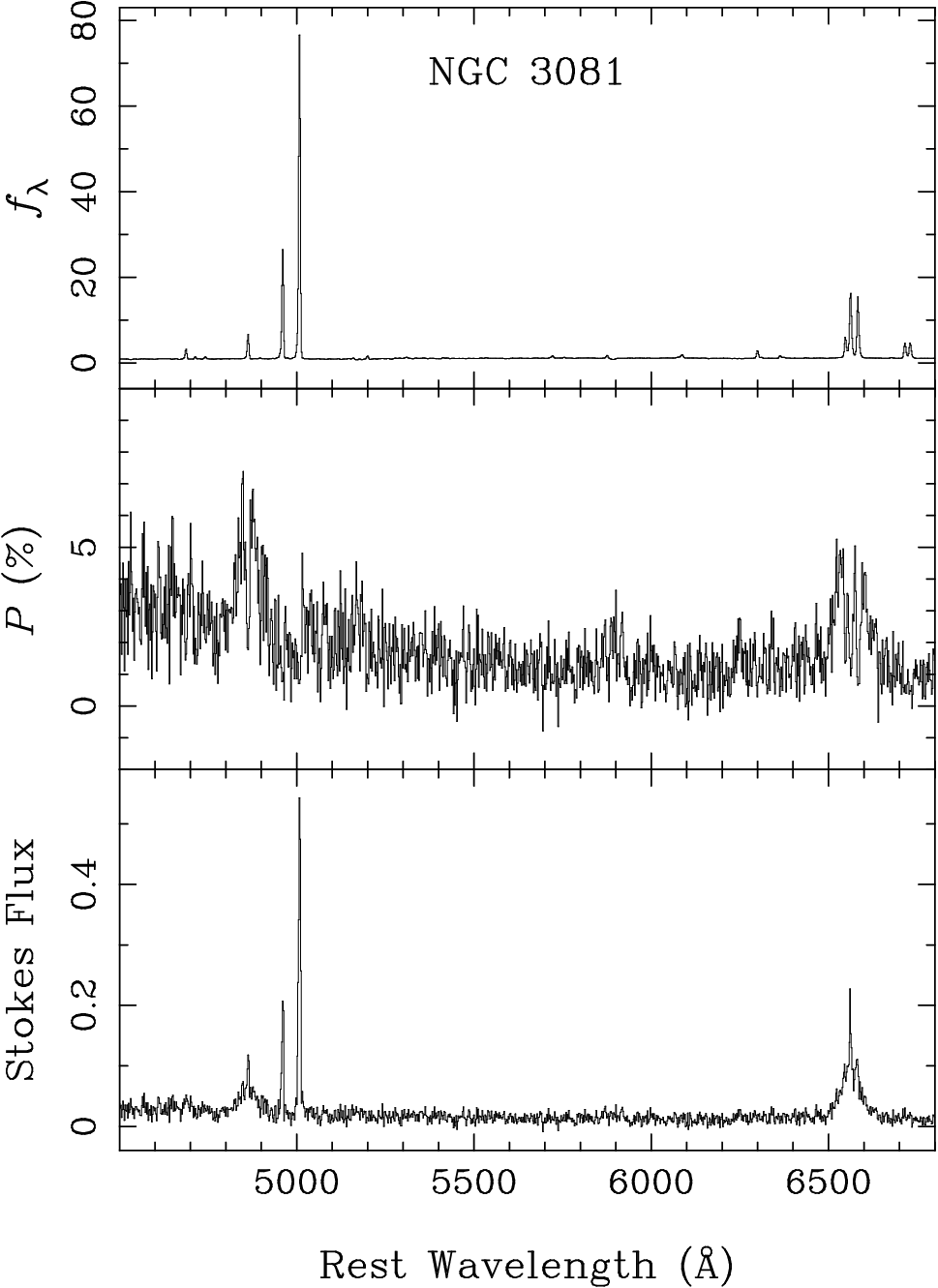}
\caption{Spectropolarimetry of NGC 3081.}
\end{figure}

\begin{figure}
\epsscale{0.7}
\plotone{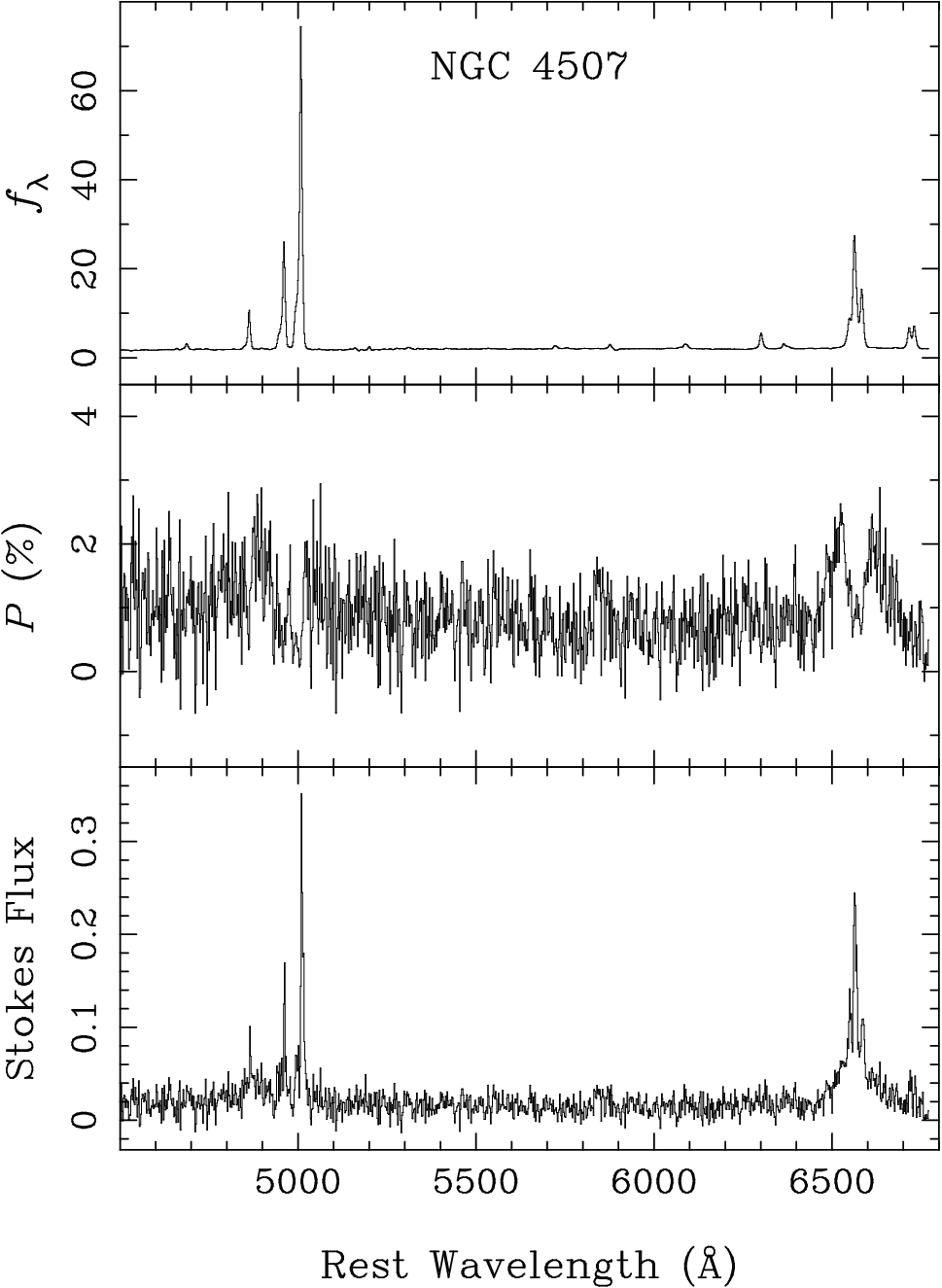}
\caption{Spectropolarimetry of NGC 4507.}
\end{figure}


\begin{thebibliography}{}

\bibitem{}Antonucci, R.~R.~J.\ 1983, Nature, 303, 158
\bibitem{}---------.\ 1984, ApJ, 278, 499
\bibitem{}---------.\ 1993, ARA\&A, 31, 473
\bibitem{}Antonucci, R.~R.~J., \& Miller, J.~S.\ 1985, ApJ, 297, 621
\bibitem{}Barth, A.~J., Filippenko, A.~V., \& Moran, E.~C.\ 1999, ApJ, 525, 673
\bibitem{}Heisler, C.~A., Lumsden, S.~L., \& Bailey, J.~A.\ 1997, Nature,
          385, 700
\bibitem{}Ho, L.~C., Filippenko, A.~V., \& Sargent, W.~L.~W.\ 1997, ApJS,
          112, 315
\bibitem{}Kay, L.~E., Magalh\~aes, A.~M., Elizalde, F., \& Rodrigues, C.\
          1999, ApJ, 518, 219
\bibitem{}Kay, L.~E., \& Moran, E.~C.\ 1998, PASP, 110, 1003
\bibitem{}Kay, L.~E., \& Moran, E.~C., Barth, A.~J., \& Filippenko, A.~V.\
          2000, in preparation
\bibitem{}Malkan, M.~A., Gorjian, V., \& Tam, R.\ 1998, ApJS, 117, 25
\bibitem{}Miller, J.~S., \& Goodrich, R.~W.\ 1990, ApJ, 355, 456
\bibitem{}Miller, J.~S., Robinson, L.~B., \& Goodrich, R.~W.\ 1988, in
          Instrumentation for Ground-Based Astronomy, ed.\ L.~B.~Robinson
          (New York: Springer), 157
\bibitem{}Moran, E.~C., Halpern, J.~P., \& Helfand, D.~J.\ 1994, ApJ, 433, L65
\bibitem{}Morganti, R., Tsvetanov, Z.~I., Gallimore, J., \& Allen, M.~G.\ 1999,
          A\&AS, 137, 457
\bibitem{}Phinney, E.~S.\ 1989, in Theory of Accretion Disks, eds.\ W.~Duschl,
          F.~Meyer, \& J.~Frank (Dordrecht: Kluwer), 457
\bibitem{}Ptak, A., Yaqoob, T., Serlemitsos, P.~J., Kunieda, H., \&
          Terashima, Y.\ 1996, ApJ, 459,~542
\bibitem{}Risaliti, G., Gilli, R., Maiolino, R., \& Salvati, M.\ 2000, A\&A,
          357, 13
\bibitem{}Risaliti, G., Maiolino, R., \& Salvati, M.\ 1999, ApJ, 522, 157
\bibitem{}Shields, J.~C., \& Filippenko, A.~V.\ 1988, ApJ, 332, L55
\bibitem{}Tran, H.~D.\ 1995a, ApJ, 440, 565
\bibitem{}---------.\ 1995b, ApJ, 440, 578
\bibitem{}Tran, H.~D., Miller, J.~S., \& Kay, L.~E.\ 1992, ApJ, 397, 452
\bibitem{}Ulvestad, J.~S., \& Wilson, A.~S.\ 1984, ApJ, 285, 439
\bibitem{}---------.\ 1989, ApJ, 343, 659 (UW89)
\bibitem{}Young, S., Hough, J.~H., Efstathiou, A., Wills, B.~J., Bailey, J.~A.,
          Ward, M.~J., \& Axon,~D.~J.\ 1996, MNRAS, 281, 1206

\end{thebibliography}
\end{document}